\documentstyle[12pt,graphicx]{article}
\newcommand{\bea}{\begin{eqnarray}}
\newcommand{\be}{\begin{equation}}
\newcommand{\eea}{\end{eqnarray}}
\newcommand{\ee}{\end{equation}}
\def\nn{\nonumber}
\def\le{\left}
\def\ri{\right}
\def\disp{\displaystyle}
\def\part{\partial}

\def\d{\delta}

\def\m{\mu}
\def\n{\nu}

\def\p{\pi}

\def\O{\Omega}

\begin{document}
%last modification  30/11/2001 15:00
\begin{center}
\vskip2cm
 {\Large\bf Semiclassical zero temperature black holes\\
  in spherically reduced theories}

\vskip2cm

{\bf C. Barbachoux}$^{a}$\footnote{Email: barba@ccr.jussieu.fr} and {\bf A.
Fabbri}$^{b}$\footnote{Email: fabbria@bo.infn.it} \\
\medskip
{\small $^{(a)}${\it LRM-CNRS, Universit\'e Pierre et Marie Curie, ERGA}} \\ {\small \it Bo\^\i te 142, 4 place Jussieu, 75005 Paris, France}\\ {\small
$^{(b)}${\it Dipartimento di Fisica dell'Universit\`a di Bologna
and INFN sezione di Bologna,}} \\ {\small {\it Via Irnerio 46,
40126 Bologna, Italy}}

\end{center}

\vskip50mm

\begin{abstract}

We numerically integrate the semiclassical equations of motion for
spherically symmetric Einstein-Maxwell theory with a
dilaton coupled scalar field and
look for zero temperature configurations. The 
solution we find is studied in detail close to the horizon 
and comparison is made with the corresponding one in the minimally coupled 
case.

\end{abstract}
\newpage
\section{Introduction}
The most attractive feature of zero temperature black holes is that 
they are the natural candidates as end-state of the
evaporation process.
 Indeed, they represent
the ideal setting where one can address the various issues connected to
the quantum evolution of the black holes, such as for instance the
problem of information loss (see e.g. \cite{infolo}).

%(for a review see for
%example~\cite{Harvey} or~\cite{Giddings}). Moreover, the
%statistical computation of the entropy of such configurations has
%been performed in string theory~\cite{Vafa} and found to be
%identical to the usual Bekenstein-Hawking
%formula~\cite{Bekenstein}\cite{Hawking}.

In spherically symmetric Einstein-Maxwell theory the only solution
with this property is the extremal Reissner-Nordstr\"om (RN) black
hole. Turning to the semiclassical theory, quantum corrections
induced by the vacuum expectation value of the stress energy tensor 
due to matter
fields  modify the spacetime geometry and it is very important to
check whether the resulting solution has still zero temperature or
not. Perturbative corrections $O(\hbar)$ to the classical geometry
evaluated close to the horizon in four dimensions do not appear to
answer unambiguously the above question \cite{AndersonI},
\cite{LoweI}. It is clear that more information would come only if
one knew the exact analytical solution to the semiclassical eqs.
of motion. For the simple case of spherically reduced
Einstein-Maxwell theory coupled with 2D minimal scalar
fields Trivedi \cite{Trivedi} was able to prove the existence of
zero temperature solutions which reduce, as $\hbar \to 0$, to the
extreme RN black hole. He also showed that although the energy
density  measured by an infalling observer close to the horizon
diverges for the classical solution, the semiclassical
configuration is regular there (only a mild singularity emerges in
the second derivative of the scalar curvature ).
 The
drawback of this analysis is that, due to the special type of
matter fields used, these results do not have an obvious four
dimensional interpretation. In order to improve this study, 
we consider here a more realistic 2D model that
recently has received a lot of attention . We employ a 2D
conformal scalar field nonminimally coupled to the dilaton field,
which classically corresponds to the s-wave sector of a 4D
minimal scalar field (this model was first studied in
\cite{model}). We  will perform a numerical integration of the
semiclassical equations of motion and show good
%resulting by considering a 2D
%conformal but nonminimally coupled scalar field which represents a
%spherically This field describes the s-wave sector of a four
%dimensional minimally coupled scalar field.
evidence that zero temperature black holes exist in this theory.
In particular , we  inspect in detail the spacetime geometry in
the region close to the horizon and compare with the results one
gets by numerical integration of the minimally coupled case.

The outline of this article is the following. In Section 2 we
briefly review the spherically reduced Einstein-Maxwell theory and
its zero-temperature solution, the extreme RN black hole. In
Section 3  the matter model we shall use will be introduced and
the expression of the $\left< T_{ab}\right>$ in  the extreme RN
background  derived.
 In Section 4 we numerically solve the backreaction equations
and  finally  section 6 contains a discussion of our
 results
 and a comparison with the case analysed in \cite{Trivedi}.

\section{Einstein-Maxwell theory in D=2}

Let us start with Einstein-Maxwell theory in four dimensions
\be
S=S_G+S_{EM}, \label{I-2} \ee where $S_G$ is the Einstein-Hilbert
action \footnote{We use units where $\hbar=G=c=k_B=1$}
\be S_G=\frac {1}{8\pi}\int d^4x \sqrt {-g^{(4)}}R^{(4)},
\label{I-3} \ee and $S_{EM}$ denotes the action associated to the
electromagnetic field
 \be S_{EM}=-\frac {1}{8\pi}\int d^4x
\sqrt {-g^{(4)}} F^2. \label{I-4} \ee  $R^{(4)}$ is the
4-dimensional scalar curvature and $F^2$ the strength of   the
electromagnetic field $F_{\m\n}$.
Assuming  spherical symmetry, the 4D metric can be written
\be
ds^2=g^{(2)}_{a\,b}dx^adx^b+e^{-2\phi(x_a)}d\O^2, \label{I-6} \ee
where $g^{(2)}_{a\,b}(x^a)$ ($a,b=1,2$) is the 2-dimensional
metric in the $(r-t)$-plane, $\phi(x_a)$ the dilaton and
$d\O^2=d\theta^2+\sin^2\theta d\phi^2$ the line element of the
unit two-sphere. Dimensional reduction of the Einstein-Hilbert
action (\ref{I-3}) can be performed by integrating over the angles
$\theta$ and $\phi$
\be
S_G^{(2)}=\frac {1}{2} \int \,d^2x \sqrt
{-g^{(2)}}e^{-2\phi}\le(R^{(2)} +2(\nabla\phi)^2+2e^{2\phi}\ri).
\label{I-7} \ee Proceeding similarly and considering $F_{\mu
\nu}=F_{\mu \nu}(x^a)$ the Maxwell action becomes
\be
S_{EM}^{(2)}=-\frac {1}{2}\int \,d^2x\sqrt
{-g^{(2)}}e^{-2\phi}\tilde F^2, \label{I-8} \ee where $\tilde F^2$
represents the field strength of a 2-dimensional gauge field. 
Black hole solutions of the theory defined by \be \label{acgrd}
S^{(2)}=S_G^{(2)}+S_{EM}^{(2)} \ee are given by the
Reissner-Nordstr\"om solution
\be
ds^2=-f(r)dt^2+\frac {1}{f(r)}dr^2,\ \ e^{-2\phi}=r^2,
 \label{I-8bc} \ee with
\be
f(r)=1-\frac {2M}{r}+\frac {Q^2}{r^2}. \label{I-8b} \ee
The parameter $M$ is the ADM mass and $Q$ the
electric charge  (the corresponding field strength is ${\tilde
F}_{rt}=\frac {Q}{r^2}$). The equation $f=0$ has two solutions for
$M>|Q|$ given by $r_{\pm}=M\pm\sqrt{M^2-Q^2}$. $r_+\equiv r_h$ and $r_-$ 
are, respectively, 
the event horizon and the inner horizon. The
Hawking temperature is
\be
T_H=\frac{ \sqrt{M^2-Q^2}}{2\pi r_h^2}.\ee Vanishing of $T_H$,
i.e. $M=|Q|$, defines the extremal configuration for which $r_+=r_-\equiv 
r_h=M$.
\section{Matter fields}
In order  to inquire on the existence of zero-temperature
solutions  in the semiclassical theory we must couple $S^{(2)}$ in
eq. (\ref{acgrd}) to quantized free matter fields. In
\cite{Trivedi} it was considered a 2D minimally coupled scalar
field described by the classical action \footnote{Usually, in
order to make physical sense of the semiclassical approximation
one considers $N$  matter fields and consider the large $N$
limit  while keeping $N\hbar$ fixed. In this way the quantum
corrections due to the other fields
  can be neglected.} \be \label{mcsc} S_M=-\frac  {1}{4}\int\,
d^2x\sqrt {-g^{(2)}} (\nabla\tilde f)^2 \label{I-u} \ee which,
after quantization, yields  the well-known Polyakov effective
action \cite{Polyakov} \be \label{pol} S_{eff}=-\frac
{1}{96\pi}\int d^2x \sqrt {-g^{(2)}} R^{(2)} \frac {1}{\Box}R^{(2)} .\ee 
This 
action
can be formally obtained by functional integration of the trace
anomaly \be \left< T\right> =\frac{R^{(2)}}{24\pi}.\ee As it was
mentioned in \cite{Trivedi}, due to the 2d origin of this
field the results one gets using $S_{tot}=S^{(2)}+S_M+S_{eff}$ do
not have an obvious four dimensional interpretation. To start
with, we shall instead consider a 4D minimally coupled scalar
field
\be
S_M^{(4)}=-\frac  {1}{16\pi}\int d^4x\sqrt {-g^{(4)}} (\nabla f)^2
\label{I-5}. \ee In a spherically symmetric spacetime, the matter
fields can be expanded into spherical harmonics, the $s$-wave
sector $\tilde f$ of the scalar field $f$ depending only on $t$
and $r$. For the $s$-wave field $\tilde f$  dimensional reduction
gives the 2D action:
\be
S_M^{(2)}=-\frac{1}{4}  \int\, d^2x\sqrt {-g^{(2)}}
e^{-2\phi}(\nabla \tilde f)^2. \label{I-9} \ee
Comparison with the scalar field in (\ref{mcsc}) shows that the
field $\tilde f$, though still 2D conformal,  has acquired a
nontrivial coupling with the dilaton field $\phi$. The
corresponding trace anomaly has additional $\phi$-dependent terms
\cite{model}
\be
\left< T\right> =\frac {1}{24\pi}\le(R^{(2)}-6(\nabla \phi)^2+6
\,\Box\phi\ri). \label{I-11} \ee
%where $\Box$ denotes the
%covariant Dalambertian.
Performing a functional integration of this expression we get the
following effective action  \cite{model}, \cite{nood}, \cite{Bal1}
\be
S^{(2)}_{eff}=-\frac {1}{2\pi}\int d^2x \sqrt {-g^{(2)}}\le[\frac
{1}{48}R^{(2)} \frac
{1}{\Box}R^{(2)}-\frac{1}{4}(\nabla\phi)^2\frac {1}{\Box}R^{(2)}
+\frac {1}{4}\phi R^{(2)}\ri]. \label{I-12} \ee  where the first
nonlocal term is the same as in (\ref{pol}) . It is important
to point out that unlike (\ref{pol}) this effective action is not
exact. Unphysical results obtained for the evaporation of
Schwarzschild black holes \cite{model}, \cite{Bal1} suggest that,
at least at finite temperature, $S^{(2)}_{eff}$ must be modified
by the addition of conformally invariant (local and nonlocal)
terms \cite{guze}, \cite{Bal3}. Considering instead
zero-temperature configurations $S^{(2)}_{eff}$ gives physically
meaningful results \cite{Bal3}. \par \noindent In conformal gauge
\be \label{coga} ds^2=-f dudv \ee this action becomes local
(i.e.$\frac{1}{\Box} R^{(2)}=-ln \, f$)  and for  static
configurations \be \label{staco} f=f(r)\ee where \be \label{nuco}
u=t-r_*,\ v=t+r_*,\ \ \  r_*=\int \frac {dr}{f(r)}\ee the
components of the 2D stress energy tensor read \bea \left<
T^{(2)}_{uu}\right> &=& \left< T^{(2)}_{vv}\right> =\frac
{1}{96\pi}\le[f\,f''-\frac {1}{2}(f')^2\ri]\nn\\
&\,&\hspace{0.9cm}+\frac {1}{64\pi}f^2\le[\le(\frac
{k'}{k}\ri)^2\,\ln\, f-\le(\frac {k'}{k}\ri)^2+2\frac
{k''}{k}\ri], \label{I-18}\\ \left< T^{(2)}_{uv}\right> &=& \frac
{1}{96\pi}f\,f''+ \frac {1}{32\pi}f\le[f'\frac {k'}{k}+f\frac
{k''}{k}- \frac {1}{2}f\le(\frac {k'}{k}\ri)^2\ri], \label{I-19}
\eea where the notation \be \label{difi} k=e^{-2\phi}\ee has been
introduced and the prime denotes derivative with respect to $r$.
Considering now the dependence on the
 dilaton field,
another relation can be deduced by functional differentiation of the
effective action (\ref{I-12}) with respect to the dilaton
\be
\frac {1}{\sqrt {-g^{(2)}}}\frac {\d S_{eff}^{(2)}}{\d \phi}=\frac
{1}{4\p}\le[\le(f'\frac {k'}{k}-f\le(\frac {k'}{k}\ri)^2+f\frac
{k''}{k}\ri)\ln\,f+f'\frac {k'}{k}-f''\ri]. \label{I-20} \ee This
term is specific to the effective action considered and does not
appear in the Polyakov theory. In a 4D viewpoint it is related to
the tangential pressure $\left<P\right>\equiv
\left<T^{\theta}_{\theta}\right>=\left< T^{\phi}_{\phi}\right>$
through the relation (\cite{model}, \cite{Bal1})
\be
\left<P\right>=\frac {1}{8\pi e^{-2\phi}\sqrt {-g^{(2)}}}\frac {\d
S^{(2)}_{eff}}{\d \phi}. \label{I-23} \ee
 For the particular case
of the extremal Reissner-Nordstr\"om black hole $f=(1-M/r)^2,
k=r^2$ we obtain the following results (see also \cite{bura}) \bea
\left< T^{(2)}_{uu}\right> &=& \left< T^{(2)}_{vv}\right> =-\frac
{1}{24\p}\frac {M}{r^3}\le(1-\frac {M}{r}\ri)^3+\frac {1}{16\p
r^2}f^2\,\ln\,f, \label{I-25b}\\ \left< T^{(2)}_{uv}\right>
&=&-\frac {1}{48\pi}\frac {M}{r^3}\le(1-\frac
{M}{r}\ri)^2\le(2-3\frac {M}{r}\ri)+\frac {1}{8\pi}\frac
{M}{r^3}\le(1-\frac {M}{r}\ri)^3, \label{I-25} \eea where the
first term on the r.h.s of these equations comes from the Polyakov
contribution to the effective action. Also, from
 Eqs. (\ref{I-20}) and (\ref{I-23})  we obtain the 4D tangential pressure:
\be
\left< P \right> =-\frac {1}{16\p^2 r^4}\le(1-\frac
{M}{r}\ri)\le(1-3\frac {M}{r}\ri)\ln\,f+\frac {M}{16\p^2
r^5}\le(4-5\frac {M}{r}\ri). \label{I-27} \ee Another important
physical quantity is
\be
F=\frac {(T^r_r-T^t_t)}{f}=\frac {4<T_{uu}>}{f^2}, \label{I-27b}
\ee  which is proportional to the energy
density measured by an infalling observer~\cite{Anderson2}.
Equation~(\ref{I-25b})
 leads to
\be
F=-\frac {1}{6\p}\frac {M}{r^2(r-M)}+\frac {1}{2\p r^2}\, \ln
\le|1-\frac {M}{r}\ri|. \label{I-26} \ee As in the minimally
coupled  case $F$ diverges when $r\to M$. The term $\sim 1/(r-M)$
is the same as that found in the Polyakov theory \cite{Trivedi},
 but despite its presence it was
shown in \cite{Trivedi} that the corresponding semiclassical
zero-temperature solution is regular at the horizon (only a mild
divergence is present in the second derivative of the scalar
curvature $R$ ). In our case , in addition to this term there
appears a subleading logarithmic divergence $\sim \ln(r-M)$, which is 
present also in the analytic approximations in four dimensions proposed in 
\cite{anapp}, as well as a nontrivial pressure $\left< P\right>$ eq.
(\ref{I-27}) . As stressed in the first of Refs. \cite{AndersonI},
the divergence of  $F$ on the horizon of the classical extreme 
black
hole makes the perturbative expansion in powers of $\hbar$ to break
down there. The calculations performed in \cite{AndersonI} are
motivated by the fact that $F$ has been proven to be finite at
$r=M$ numerically in D=4 \cite{Anderson2}, but due to the result
(\ref{I-26})  reliable near horizon calculations 
for zero temperature black holes
performed using the
effective action (\ref{I-12}) must be nonperturbative in $\hbar$. 
Similarly, the $O(\hbar)$ results presented in
\cite{medved} (obtained by considering near-extreme black holes in the 
near horizon region) do not appear to have much  physical meaning.

\section{Backreaction}

We now come to the main question addressed in this paper: do
self-consistent zero-temperature black holes exist in the
semiclassical theory ? For this purpose,  we need first of all to
write down the semiclassical  Einstein equations, which can be
derived by differentiation of the action
$S_{tot}^{(2)}=S^{(2)}+S_M^{(2)}+S_{eff}^{(2)}$ (see eqs.
(\ref{acgrd}), (\ref{I-9}) and (\ref{I-12}) ) with respect to the
2d metric $g_{ab}^{(2)}$ and the dilaton field $\phi$. In
conformal gauge (\ref{coga}) and considering static configurations
(\ref{staco}), (\ref{nuco}) the relevant expressions concerning
the matter part of the action have been derived in
  (\ref{I-18}), (\ref{I-19}) and (\ref{I-20}).
The corresponding quantities coming from the gravity and
electromagnetic actions (where $F^2=-2\frac {Q^2}{k^2}$) can be
easily obtained by differentiation of (\ref{I-7}) and (\ref{I-8}).

The $uu$ (or $vv$) constraint reads
\bea 0&=&f^2k''-\frac {1}{2}\le(f\frac
{k'}{k}\ri)^2k+\xi\le(f''f-\frac {1}{2}(f')^2\ri)\nn\\
&\,&\hspace{1cm}+3\xi\le[\frac {1}{2}\le(f\frac {k'}{k}\ri)^2ln\,f
-\frac {1}{2}\le(f\frac {k'}{k}\ri)^2+3f^2\frac {k''}{k}\ri],
\label{II-5} \eea where the coefficient $\xi=\disp\frac
{\hbar}{12\pi}$ has been introduced (we have reintroduced $\hbar$
in the formulas in order to make the distinction between classical
and quantum terms more clear) . The equation obtained by varying
the trace of the metric (i.e. $g_{uv}$) reads
\be
0=-2+f'k'+fk''+2\frac {Q^2}{k}+\xi
f''+3\xi\le[f'\frac{k'}{k}+f\frac {k''}{k} -\frac {1}{2}f\le(\frac
{k'}{k}\ri)^2\ri] \label{II-6}. \ee
Finally, differentiation with respect to $\phi$ gives
\bea 0=f''&-&\frac {1}{2}f\le(\frac {k'}{k}\ri)^2+f'\frac
{k'}{k}+f\frac {k''}{k} -\frac {2Q^2}{k^2}\nn\\ &-&3\xi\le[\frac
{f'k'}{k^2}-\frac {f''}{k}+ ln\,f \le(\frac {f'k'}{k^2}-f\frac
{(k')^2}{k^3}+f\frac {k''}{k^2}\ri)\ri]. \label{II-7} \eea
For $\xi=0$ (\ref{II-5}), (\ref{II-6}) and (\ref{II-7}) are the
classical eqs. of motion , for which the only zero-temperature
configuration is the  extremal Reissner-Nordstrom black hole
$f=(1-M/r)^2,\ k=r^2$. In the quantum terms, we have separated the
ones multiplying $\xi$, coming to the Polyakov contribution to
the effective action and present also in \cite{Trivedi}, and those 
proportional to $3\xi$ representing the
additional contributions in the effective action $S_{eff}^{(2)}$
(Eq. (\ref{I-12})).

As the three previous equations involve only two independent
functions $f(r)$ and $k(r)$, one is redundant. Indeed, they are
 related through the Bianchi identities combined with the
 ``nonconservation'' eqs. for the matter part \cite{Bal1}
\be
\nabla_{a}\left< T^{(2)\,a}_b\right> +8\pi e^{-2\phi}\left<
P\right> \nabla_b\phi=0. \label{II-8} \ee 
%Therefore the
%determination of the solutions of these equations requires only
%two of these equations.
\par \noindent
Also, as these non-linear differential equations involve the
second
 order derivatives of $f$ and $k$, two boundary conditions on
 these functions are required to determine them uniquely. 
For a 
 zero temperature black hole, natural boundary conditions can be
 imposed on the function $f$ at the horizon. First of all, $f$ has to 
vanish
 there. Moreover, in the gauge used the temperature of the black
 hole takes the simple expression
\be
T_H=\frac{ \kappa}{2\pi},  \label{I-17} \ee where
\be
\kappa=\frac {1}{2} \le. f'\ri|_{r=r_h} \label{I-16} \ee is the
surface gravity and $r_h$ denotes the radius of the horizon.
So $T_H=0$ means $f'=0$ at $r=r_h$.
\par \noindent
In the Polyakov case, starting from these boundary conditions
Trivedi~\cite{Trivedi} has found the form of the exact solution
(nonperturbative in $\hbar$) close to the horizon as a
(non-analytic) expansion in powers of the coordinate distance from
the horizon $ r-r_h$. In our case, however, the terms proportional
to $ln\,f$ complicate exceedingly this analysis and seem to
prevent an expansion in closed form of the solution for small
values of $r-r_h$ analogous to that proposed in \cite{Trivedi}.

A numerical resolution has then been undertaken and the boundary
conditions have been imposed at infinity \cite{Lowe}, where the
solution is to a good approximation the extreme RN. In this region
apart from a finite but very small renormalization of the
classical mass (following \cite{LoweI} it is 
$M_R/Q=1+O(\frac{\xi^2}{Q^4})$) the
first quantum corrections  to the spacetime metric are of the order 
$O(1/r^3)$.
\par \noindent
To start with, we have introduced dimensionless variables and
functions in the differential equations that we have to integrate:
$$x=r/Q,\;\;\;\tilde {k}(x)=k(r)/Q^2,\;\;\;\tilde f(x)=f(r).$$
This means choosing the black hole charge to be the natural
unit of length.
\par \noindent
Our numerical integrations have been performed using the
$uu$-constraint equation (\ref{II-5}) and the $\phi$-equation
(\ref{II-7}) for the value $\disp\frac {\xi}{Q^2}=10^{-5}$. All
along our calculations, the solutions of these equations have been
checked to be compatible with eq. (\ref{II-6}) as well with a
precision less than $10^{-7}$.
\par \noindent
In order to probe the accuracy of our numerical simulations, we
have first considered the integration of the semiclassical eqs.
for the minimally coupled case  (i.e. discarding the
terms proportional to $3\xi$ in Equations~(\ref{II-5}),
(\ref{II-6}) and~(\ref{II-7})) and compared the numerical results
with the form of the exact solution close to the horizon given
 by Trivedi~\cite{Trivedi}. We find that,
for a value of the horizon $x_P=1,0229$ (in units where $Q=1$),
i.e. with a deviation of about $2,3\%$ from the classical value,
the functions $f$ and $f'$ behave like those in  \cite{Trivedi}
with a precision of about $5 \cdot 10^{-4}$
 when $x\to x_P$ and that $k$ is accurate
  with a precision $5 \cdot 10^{-5}$ and $k'$ with the accuracy $2 \cdot 
10^{-4}$.
  We can then expect the global precision of our simulation to be at
  least about $2 \cdot 10^{-2}\%$.
\par
\begin{figure}[ht]
\centerline{
\hspace{-1cm}
\includegraphics[scale=0.8]{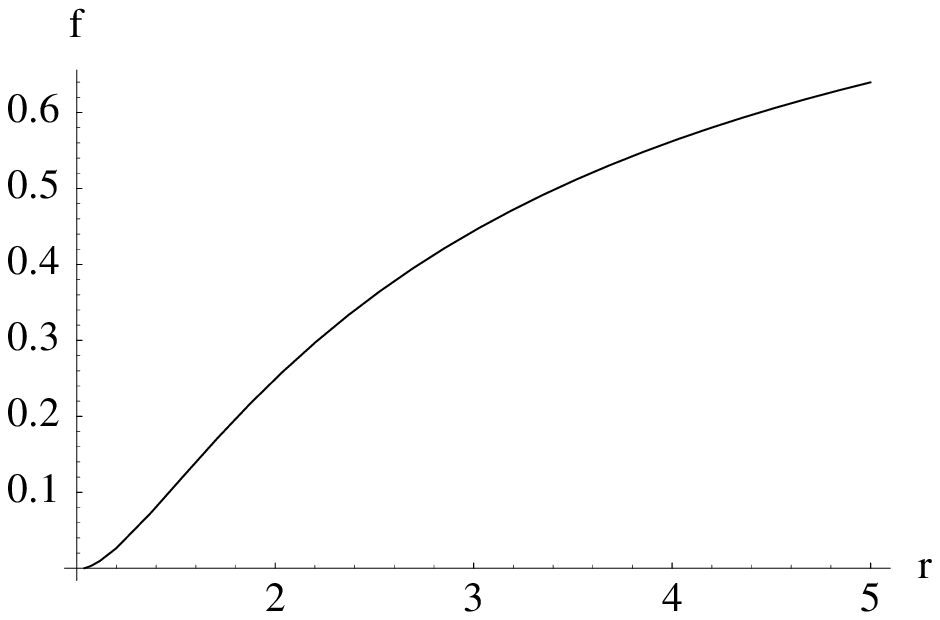},\includegraphics[scale=0.8]{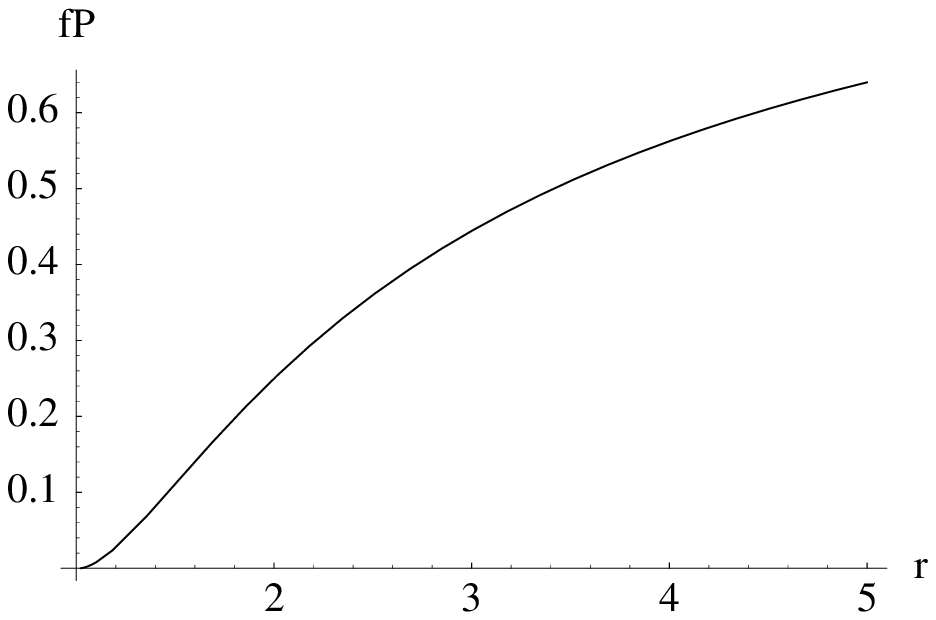}}
\caption{Plot of the function $f$ for our model (left) and the
corresponding $f$ for the minimally coupled case (right).  We have set  
$Q=1$.
}
\label{g1}
\end{figure}

\begin{figure}[ht]
\centerline{
\hspace{-1cm}\includegraphics[scale=0.8]{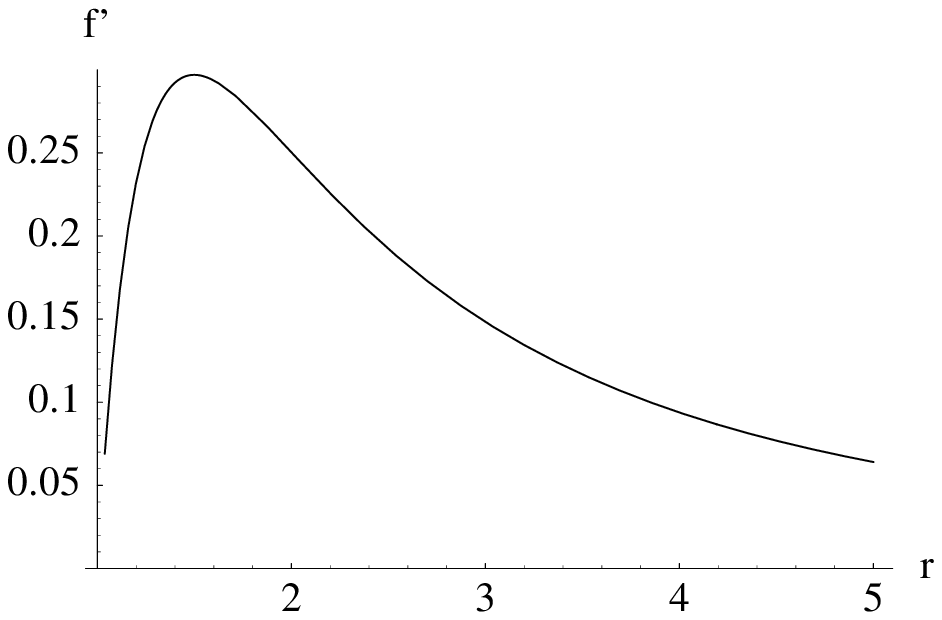},\includegraphics[scale=0.8]{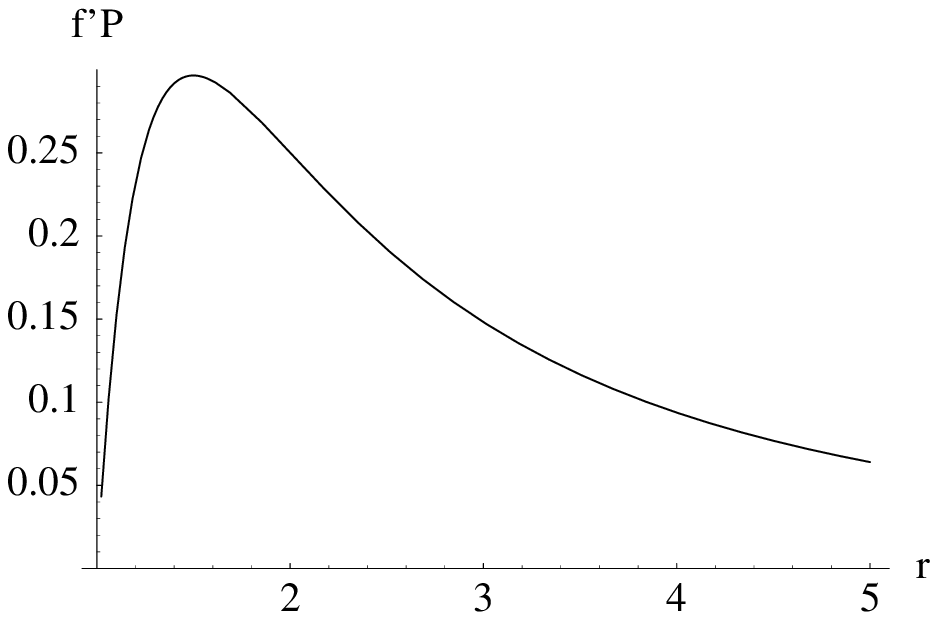}}
\caption{Comparison of the values of $f'$ for the two
theories.} \label{g2}
\end{figure}

\begin{figure}[ht]
\centerline{\hspace{-1cm}\includegraphics[scale=0.8]{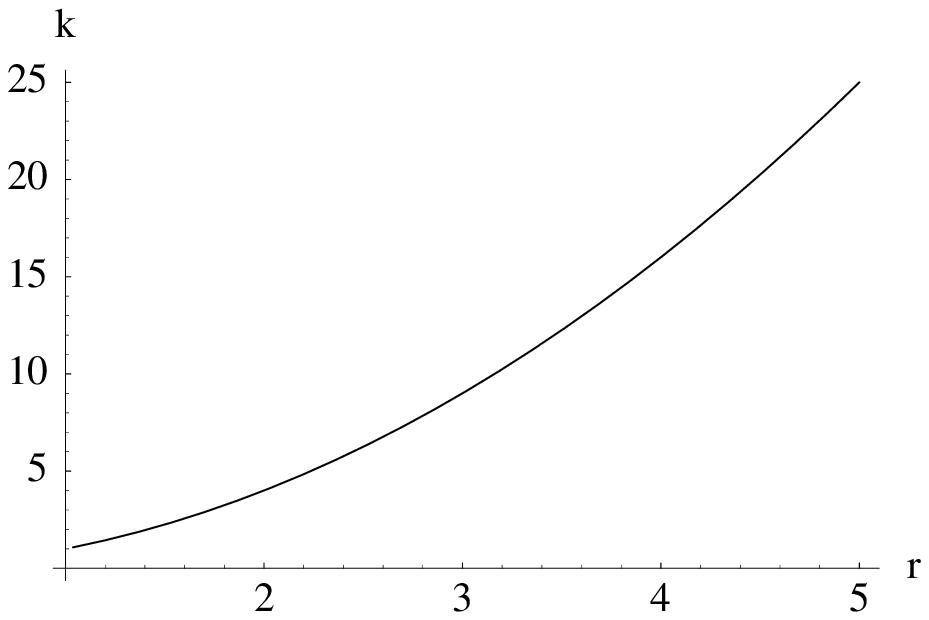},\includegraphics[scale=0.8]{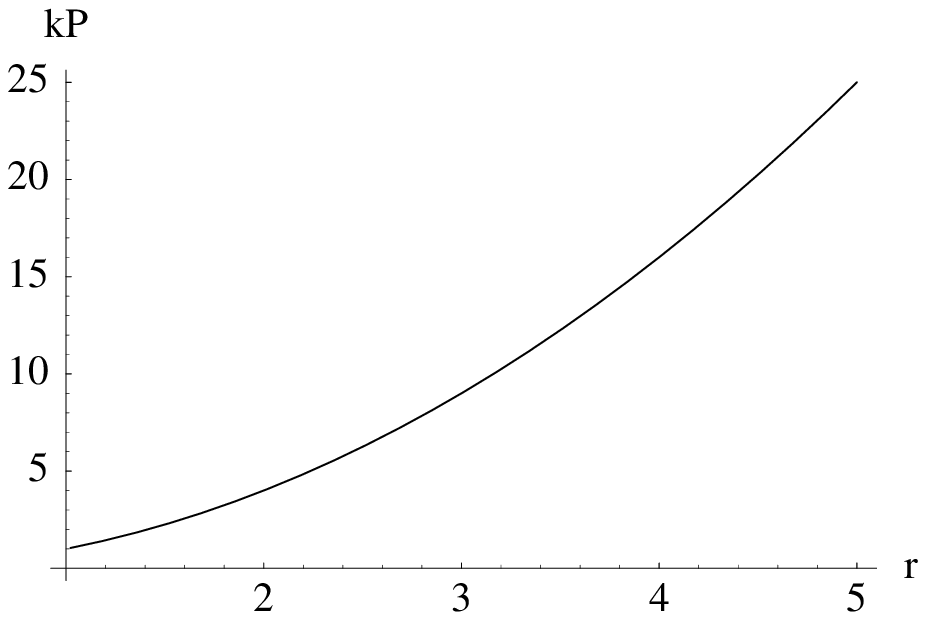}}
\caption{Plots of the function $k$. } \label{g3}
\end{figure}

\begin{figure}[here!]
\centerline{\hspace{-1cm}\includegraphics[scale=0.8]{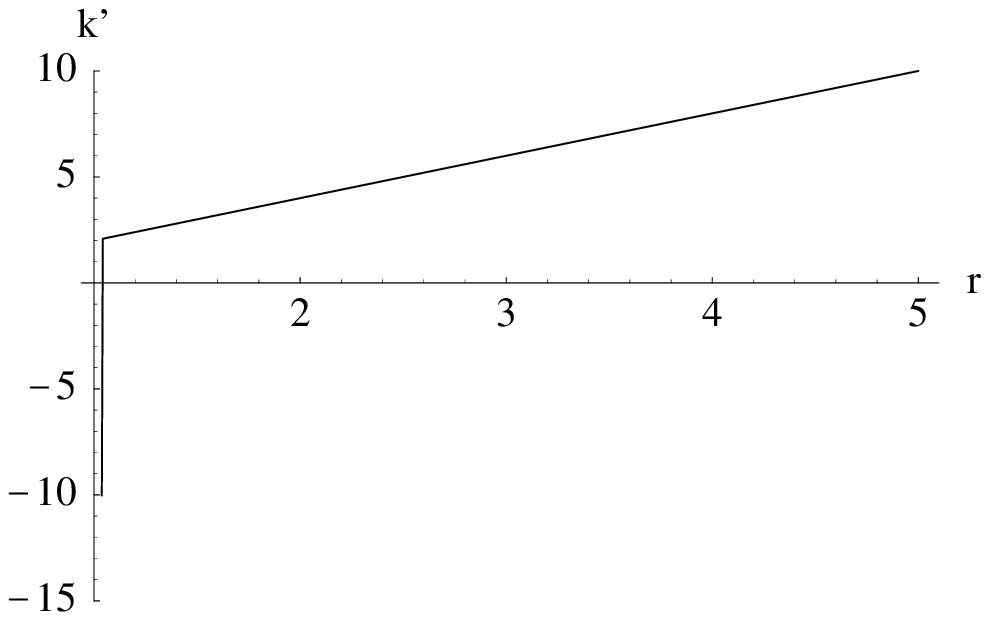},\includegraphics[scale=0.8]{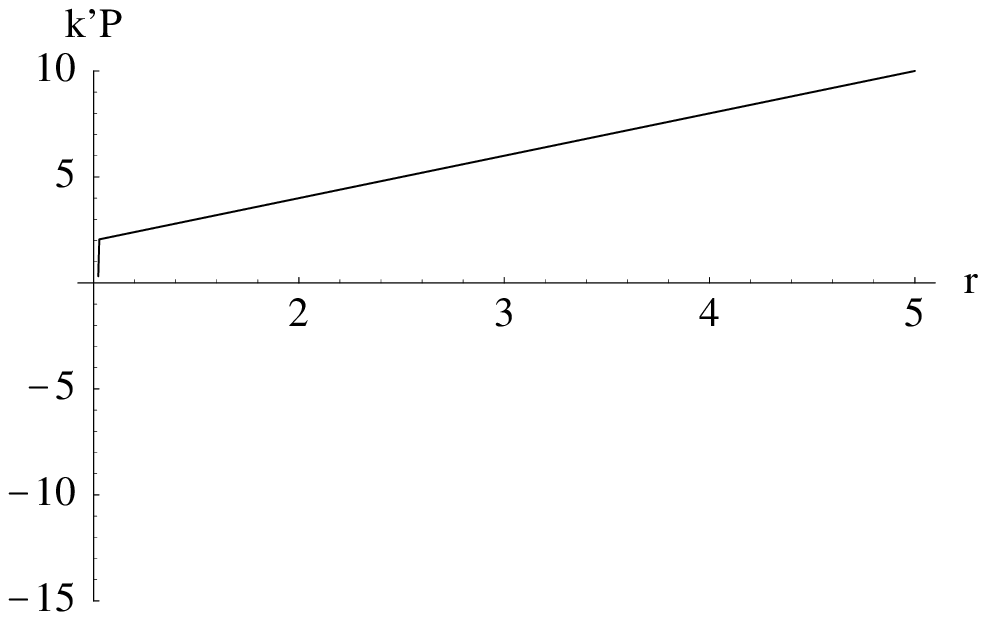}}
\caption{Plots of the functions $k'$. } \label{g4}
\end{figure}

 \noindent
Considering now the nonminimally coupled case, we have integrated
Equations~(\ref{II-5}) and~(\ref{II-7}). The results of this
simulation are illustrated by the plots on the left of  
Figs.\ref{g1}-\ref{g4}
where the functions $f$ and $k$ and their first derivatives have
been shown for $x$ varying from the horizon $x_D=1,0378$ to $5$
(in units where $Q=1$). To facilitate the comparison, the same
functions in the minimal case (the plots on the right of  
Figs. \ref{g1}-\ref{g4}) have been reported for $x$ varying
from $x_P=1,0229$ to $5$.

\section{Discussion and conclusions}

Our numerical simulations presented in Figs. 1-4 and the
comparison with the corresponding solution of the Polyakov theory
appear to give good evidence that zero-temperature configurations
exist
in this theory.
To get some insights from these results, we have compared the
numerical values close to the horizon of these solutions with
those obtained in the minimally coupled case. It turns out that
the differences between the values of the functions $f$ and $f'$
for the two theories are less than $8 \cdot 10^{-4}$ (as deduced 
previously,  
the numerical
precision is about $2 \cdot 10^{-4}$ ). The same
reasoning applies to the function $k$ with a difference less than
$8 \cdot 10^{-5}$. The first noticeable difference between the two
theories appears at the first order derivative of the function
$k$: the value of $k'$ on the horizon for our model is estimated
at $-18,094$ compared to $+2,075$ for the Polyakov case. Going
further in the derivatives of the function $f$ we have that the
third derivative of $f$ blows up at the horizon for the
nonminimally coupled field compared to the divergence of the
fourth derivative in the minimal case. Moreover, it is interesting
to stress that the value of the coordinate $x$ at the horizon
$x_D=1,0378$ (in units where $Q=1$) differs from $x_P=1,0229$ of
about $1,5\%$ (up to a numerical precision of about $2 \cdot 
10^{-2}\%$).\\ 
\noindent
Curvature invariants can be easily calculated starting from the
results presented here. The 2D Ricci scalar $R^{(2)}=-f''$ is
finite on the horizon , as it is shown in the plot of Fig. 5.
Moreover, finiteness of $k$ and $k'$ are enough to prove that also
the corresponding four dimensional scalar curvature $R^{(4)}$ is regular 
as the horizon 
is approached. A similar
conclusion, despite the divergence of $F$ at the horizon of the
extreme RN black hole, was found 
in the minimally coupled case \cite{Trivedi}.
%despite the divergence of the energy density as measured by an
%infalling observer in the classical extreme RN black hole. 
In our case, since
the leading term in eq. (\ref{I-26}) is the same as in the
Polyakov theory, it is reasonable that the main conclusion about
the regularity of the geometry at the horizon is unchanged.
The difference with respect to the case analysed in \cite{Trivedi} is that 
the
``mild'' singularity appearing at the horizon in the second
derivative of the curvature 
close to the horizon ($\sim f''''$) is replaced by a ``stronger''
divergence in its first derivative (i.e. $f'''$) . This is due to
the logarithmic divergence $\ln (f)$ appearing in eq. (\ref{I-26})
as well as in eqs. (\ref{II-5}) and (\ref{II-7}).

In conclusion, by  numerical integration of the two-dimensional
semiclassical equations of motion for the case of spherically reduced
Einstein-Maxwell theory and a scalar field nonminimally
coupled to the dilaton we have found solutions describing
zero-temperature black holes.
 Similarities and differences
with respect to the simpler minimally coupled case
have been studied in detail. Due to the intrinsic four dimensional
nature of the matter field used, our results could be relevant in
order to address the same issue in the physical world $D=4$ .
Finally, following Refs. \cite{AndersonI} an interesting extension
of this work would be to check whether the solution found here
does indeed represent the end-point of the evaporation process
(for the minimally coupled case this problem has been addressed 
both in the near-horizon approximation  \cite{fadijo} and in 
the whole spacetime numerically  \cite{sopi}) .

\section*{Acknowledgements}
We would like to thank R. Balbinot and P. Nicolini for useful
discussions and for collaboration at an earlier stage of this
work. C.B. was supported by the Italian Minister of Foreign Affairs.

\newpage
\thispagestyle{empty}
\vspace{2cm}
\newpage
\thispagestyle{empty}
\vspace{2cm}
\newpage
\thispagestyle{empty}
\begin{figure}[here!]
\centerline{\hspace{-1cm}\includegraphics[scale=0.8]{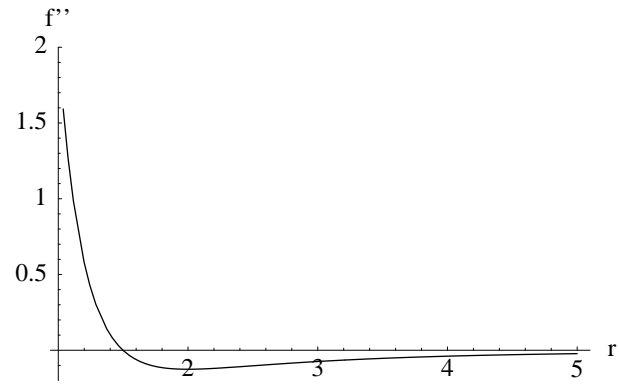}}
\caption{Values of
$f''=-R^{(2)}$   for our model.} \label{g5}
\end{figure}

\end{document}